\documentclass[oneside,final,11pt]{amsart}

\usepackage{hyperref}
\usepackage[font=small]{caption}
\usepackage{graphicx}
\usepackage{dsfont}

\newcommand{\R}{\mathds R}

\DeclareMathOperator{\Binom}{Binom}

\title[Incorrect use of Carlisle's method for dichotomous variables]{On the incorrect use of Carlisle's method for dichotomous variables}

\author{Daniel V. Tausk}
\address{Departamento de Matem\'atica,\hfill\break\indent Universidade de S\~ao Paulo, Brazil}
\email{tausk@ime.usp.br}
\urladdr{http://www.ime.usp.br/\~{}tausk}

\date{August 31st, 2022}

\begin{document}

\theoremstyle{remark}\newtheorem{rem}{Remark}

\begin{abstract}
In 2017, J. B. Carlisle has proposed a method for fraud detection in randomized controlled trials based on a comparison of reported baseline data between treatment groups. While Carlisle has only used the method for continuous variables, some authors have recently employed a naive adaption of the method for dichotomous variables. We explain why such adaptation leads to p-values that are wrong by orders of magnitude and we make a simple concrete proposal for correction of the method.
\end{abstract}

\maketitle

\begin{section}{Introduction}

The fraud detection method proposed by Carlisle in \cite{Carlisle} consists of the following steps: (i) for each continuous variable in the reported baseline data of a randomized controlled trial, compare treatment and control groups and compute a p-value (using, for instance, a t-test); (ii) combine all the p-values obtained in the first step (using, for instance, Stouffer's method) into a single p-value. If the combined p-value obtained from this procedure is either too small or too close to $1$, we flag the study as ``odd'' or ``suspicious''.

A very small combined p-value corresponds to baseline data that is too unbalanced and a combined p-value that is very close to $1$ corresponds to baseline data that is oddly overly balanced, what could be informally described as ``too good to be true''. The rational behind the method is that we are testing the null hypothesis that the baseline data arose from a well-performed randomization procedure against some alternative hypothesis (not rigorously defined in mathematical terms) that the data was generated from poor methodology or was fabricated. It is implicitly assumed that poor methodology could increase the probability of unbalanced baseline data, leading to small p-values, and data fabrication by hand (without the help of automatic pseudorandom number generators) could increase the probability of overly balanced baseline data and hence p-values that are close to $1$.

Carlisle's method is known to have several limitations (see, for instance, \cite{Senn,Wilson}), some of which are recognized by Carlisle himself \cite{Carlisle}:
\begin{itemize}
\item the methods used for combination of p-values assume independence, while baseline variables in clinical trials are usually not independent;
\item stratified randomization leads to more similar groups than standard simple randomization and thus p-values computed without taking stratification into account will tend to be closer to $1$;
\item the method often detects simple mistakes and typos and in fact it doesn't seem to have much power to detect actual fraud. It should thus be regarded as a (not very good) screening test, not a diagnostic test.
\end{itemize}
To this list I would add the issue of multiple testing: fraud detectives often perform multiple statistical tests on multiple articles and thus some small p-values (or some p-values too close to $1$) are supposed to be found among honest methodologically sound trials. A cutoff such as $0.05$ or $0.01$ is thus not enough evidence to make an accusation. Carlisle \cite{Carlisle} implicitly suggests a threshold of $1$ in $10{,}000$, unless several odd p-values are found among studies by the same authors.

Although Carlisle restricts his analysis to continuous baseline variables, some authors have incorrectly adapted the method to dichotomous variables, applying Fisher's exact test or chi-square test with Yates continuity correction for the computation of individual p-values. This leads to gross errors for reasons explained in Sections~\ref{sec:pitfalls} and \ref{sec:combining}. In Section~\ref{sec:correctmethod} we suggest a corrected adaptation of Carlisle's method to dichotomous variables. Finally, in Sections~\ref{sec:Sheldrick} and \ref{sec:Brown} we look at a few concrete examples in which the incorrect adaptation of Carlisle's method to dichotomous variables has been applied.

All computations for this manuscript were made using R software \cite{R} and all code is available at \href{http://www.ime.usp.br/~tausk/Carlisle}{http://www.ime.usp.br/\~{}tausk/Carlisle}. Numbers presented in decimal notation are often rounded to a few decimal places without explicit mention.

\end{section}

\begin{section}{The pitfalls of statistics involving small numbers}
\label{sec:pitfalls}

Let us start by briefly recalling some of the theory of p-values and combination of p-values. Since this is all well-known by most statisticians, I will keep technical details to a minimum to make the text accessible to a wider audience.

Recall that a p-value is a $[0,1]$-valued statistic, i.e., a function of the outcome of some random experiment taking values between $0$ and $1$. The p-value is used as a decision criterion for rejecting some null hypothesis: if one chooses a number $\alpha\in[0,1]$ and decides to reject the null hypothesis when the p-value $\mathfrak p$ is less than or equal to $\alpha$, then the probability of committing a type I error (i.e., incorrectly rejecting a true null hypothesis) should ideally be equal to $\alpha$. In symbols, ideally we should have:
\begin{equation}\label{eq:puniform}
\mathbb P_0(\mathfrak p\le\alpha)=\alpha,
\end{equation}
for all $\alpha\in[0,1]$, where $\mathbb P_0$ denotes the probability of an event under the null hypothesis. The number $\alpha$ is often called the {\em nominal significance level\/} of the test and the probability $\mathbb P_0(\mathfrak p\le\alpha)$ is the true probability of type I error. Equality \eqref{eq:puniform} means that $\mathfrak p$ is a random variable whose distribution is uniform in the interval $[0,1]$. Such uniform distribution would be empirically observed if many independent random experiments were performed and the corresponding p-values were plotted.

Unfortunately, equality \eqref{eq:puniform} between true probability of type I error and nominal significance level is not always valid, for reasons that we explain in a moment. If instead of \eqref{eq:puniform} we have the inequality
\begin{equation}\label{eq:pbelowuniform}
\mathbb P_0(\mathfrak p\le\alpha)\le\alpha,
\end{equation}
for all $\alpha\in[0,1]$ then the test is {\em conservative}: the probability of committing a type I error can be smaller than the chosen nominal significance level $\alpha$, but will never exceed it. In this case type I errors will not occur more often than what we choose to be an acceptable error rate, but a conservative test might have {\em low power\/}, i.e., false null hypotheses will frequently not be rejected. On the other hand, if inequality \eqref{eq:pbelowuniform} can fail then we may be committing a type I error more often than we wish.

The standard way of obtaining a p-value is the following: one chooses a {\em test statistic\/} $T$ with the property that, say, smaller values of $T$ correspond to observed outcomes that are less compatible with the null hypothesis and more compatible with the alternative hypothesis. If a certain value $t$ of $T$ is obtained in the experiment, we compute the corresponding p-value as the probability $\mathbb P_0(T\le t)$ under the null hypothesis that the value of $T$ be less than or equal to $t$. It is easy to prove that a p-value defined in this manner will always satisfy\footnote{%
Sometimes the probability $\mathbb P_0(T\le t)$ is replaced by a conditional probability of the form $\mathbb P_0(T\le t|S=s)$, with $S$ some random object whose observed value is $s$. This type of conditioning is often useful for eliminating unknown {\em nuisance parameters}, i.e., parameters whose value is not fixed by the null hypothesis. Inequality \eqref{eq:pbelowuniform} also holds when a conditional probability is used to define the p-value.\label{foot:conditional}}
inequality \eqref{eq:pbelowuniform}, but equality \eqref{eq:puniform} will not hold unless the random variable $T$ is {\em continuous}, i.e., $\mathbb P_0(T=t)=0$ for all $t$.

We note that in many cases even inequality \eqref{eq:pbelowuniform} can fail due to the fact that one is using approximations to the true distribution of $T$ under the null hypothesis. For example, one might be using the Central Limit Theorem to approximate the distribution of a large sample mean by a normal distribution or one might be replacing the value of an unknown parameter by a point estimate (say, replacing unknown population variance by sample variance). While most statistics users know that equality \eqref{eq:puniform} and even inequality \eqref{eq:pbelowuniform} will fail to hold exactly when approximations are employed, some might fail to notice that substantial deviations from equality \eqref{eq:puniform} can occur due to failure of continuity of the test statistic even when so called ``exact'' tests are used. For example, if the test statistic $T$ can assume only a finite number of values then there will also be only a finite number of possible p-values and the probability $\mathbb P_0(\mathfrak p\le\alpha)$ will be equal to the largest possible p-value that is less than or equal to $\alpha$; this might be a lot smaller than $\alpha$.

Let us look at the concrete example which is most relevant for our purposes. Recall that a $2\times2$ {\em contingency table\/} (see Figure~\ref{fig:contingency}) is used to summarize the number of patients in each treatment group for which a certain dichotomous variable has the value ``yes'' and the number of patients in each treatment group for which such dichotomous variable has the value ``no''.

\begin{figure}
\begin{tabular}{|c|c|c|c|}
\hline
\rule{0pt}{15pt}&group 1&group 2&row total\\
\hline
\rule{0pt}{15pt}yes&$k_1$&$k_2$&$s$\\
\hline
\rule{0pt}{15pt}no&$N_1-k_1$&$N_2-k_2$&$N-s$\\
\hline
\rule{0pt}{15pt}column total&$N_1$&$N_2$&$N$\\
\hline
\end{tabular}
\caption{A $2\times2$ contingency table. First group has $N_1$ patients of which $k_1$ are ``yes'' patients and second group
has $N_2$ patients of which $k_2$ are ``yes'' patients; $s=k_1+k_2$ is the total number of ``yes'' patients and $N=N_1+N_2$ is the total number of patients.}\label{fig:contingency}
\centering
\end{figure}

{\em Fisher's exact test\/} can be applied to such a contingency table to test the null hypothesis that the elements $k_1$ and $k_2$ on the first row were independently sampled from binomial distributions $\Binom(N_1,p)$ and $\Binom(N_2,p)$ with the same probability parameter $p$. Such null hypothesis will be true if the value of the dichotomous variable is independent among patients and if the probability $p$ of a value ``yes'' is the same for both treatment groups. The test works by conditioning on the row total $s$, which eliminates the dependence on the unknown parameter $p$ and makes the distribution of $k_1$ a hypergeometric distribution with all parameters known; moreover, $k_2=s-k_1$ is then determined by $k_1$. The test statistic $T$ used to rank the contingency tables in terms of compatibility with the null hypothesis is often chosen as the hypergeometric probability of $k_1$ and then a p-value is obtained in the usual way. Note that for a given value of $s$ the test statistic $T$ and thus the corresponding p-value can assume at most $s+1$ distinct values, as $k_1$ must be an integer between $0$ and $s$. Thus, for instance, in situations where $pN$ is small, the value of $s$ will have a tendency to be small and substantial deviations from equality \eqref{eq:puniform} will occur.

Let us illustrate the problem with some concrete numbers. Assume that we have two groups of $100$ patients each and that the probability of a patient being a ``yes'' patient is $0.05$, i.e., the elements of the first row of the contingency table are independently sampled from $\Binom(100,0.05)$. There are $101^2=10{,}201$ possible contingency tables and the distribution of Fisher's exact p-value can be determined explicitly by calculating both the p-value and the probability of each possible contingency table. In Figure~\ref{fig:Fisher} we plot the cumulative distribution function of Fisher's exact p-value (in red) and the cumulative distribution function of a uniform distribution (black diagonal). The graph of the cumulative distribution function of a p-value can be thought as a graph in which the horizontal axis is the nominal significance level and the vertical axis is the true probability of type I error. The diagonal corresponds to the ideal case \eqref{eq:puniform} of a uniformly distributed p-value. If only inequality \eqref{eq:pbelowuniform} holds then the line stays below the diagonal and if even inequality \eqref{eq:pbelowuniform} fails then the line can go above the diagonal.

\begin{figure}
\includegraphics[scale=0.6]{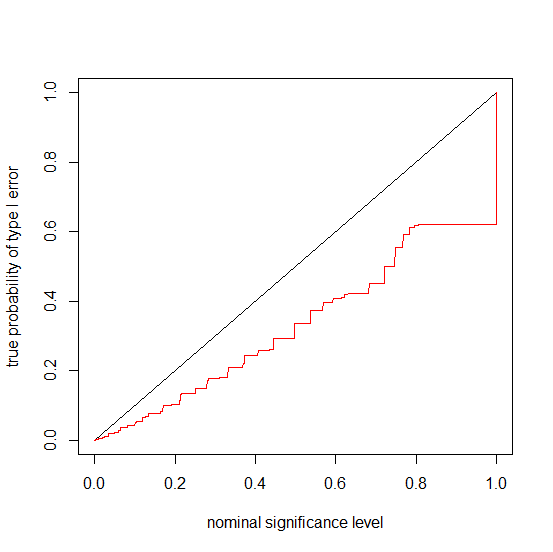}
\caption{Cumulative distribution function of Fisher's exact p-value (red) versus uniform distribution (black). $100$ patients per group,
probability of ``yes'' equal to $0.05$.}\label{fig:Fisher}
\centering
\end{figure}

As the graph illustrates, there is in this case a substantial deviation from a uniform distribution. One can also compare the distribution of the p-value with a uniform distribution by looking at the expected value. For a uniform distribution, the expected value is of course $\frac12$, while for Fisher's exact p-value in this example the expected value is $0.6699$.

Another popular test for the same null hypothesis of Fisher's exact test is the chi-square test. It uses a statistic whose distribution under the null hypothesis is approximately a chi-square distribution and it is not an exact test. For $2\times2$ contingency tables, the statistical software R uses by default the so called Yates continuity correction for the chi-square statistic. In Figure~\ref{fig:Fisherchiyates} we compare the cumulative distribution functions of Fisher's exact p-value (red), chi-square test p-value (green) and chi-square test with Yates continuity correction p-value (blue) again in the case of $100$ patients per group and a $0.05$ probability of ``yes''. While the green line sometimes goes above the diagonal, illustrating the fact that
inequality \eqref{eq:pbelowuniform} sometimes fails, it is typically much closer to the diagonal than the red and blue lines. The expected value of the chi-square test p-value is $0.4997$. On the other hand, the blue line is almost identical to the red line (making the red line barely visible), which means that once Yates continuity correction is applied to the chi-square test the distribution of the p-value under the null hypothesis is almost identical to the distribution of Fisher's exact p-value in this example. The expected value of the p-value for chi-square test with Yates continuity correction is $0.6694$.

\begin{figure}
\includegraphics[scale=0.6]{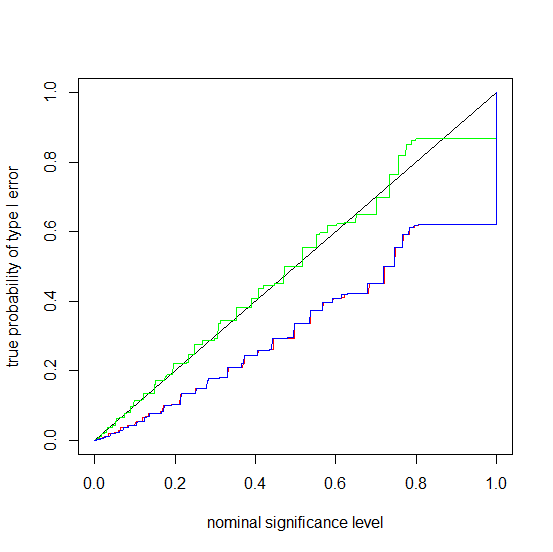}
\caption{Cumulative distribution function of Fisher's exact p-value (red), chi-square test p-value (green) and chi-square test with Yates continuity correction p-value (blue) versus uniform distribution (black). $100$ patients per group,
probability of ``yes'' equal to $0.05$.}\label{fig:Fisherchiyates}
\centering
\end{figure}

But so what if the distribution of the p-value of Fisher's exact test (or chi-square test with Yates continuity correction) can deviate a lot from a uniform distribution? One might be interested in looking for more powerful options, but as long as inequality \eqref{eq:pbelowuniform} is valid, we can at least confidently reject the null hypothesis when the p-value is very small. However, in the context of Carlisle's method we are sometimes going to reject the null hypothesis when the p-value $\mathfrak p$ is very close to $1$ or, equivalently, we are going to regard $1-\mathfrak p$ as the p-value of our test, rejecting the null hypothesis when $1-\mathfrak p$ is very small. Unfortunately, if $\mathfrak p$ satisfies inequality \eqref{eq:pbelowuniform} then $1-\mathfrak p$ satisfies the reverse inequality:
\[\mathbb P_0(1-\mathfrak p\le\alpha)=\mathbb P_0(\mathfrak p\ge1-\alpha)=1-\mathbb P_0(\mathfrak p<1-\alpha)\ge1-\mathbb P_0(\mathfrak p\le1-\alpha)\ge\alpha,\]
so that now the true probability of type I error will be above the nominal significance level. In case the distribution of $\mathfrak p$ deviates substantially from a uniform distribution, the true probability of type I error can be much larger than nominal significance level when $1-\mathfrak p$ is used as p-value and this is really inappropriate. In Figure~\ref{fig:oneminusFisher} we compare the cumulative distribution function of $1$ minus Fisher's exact p-value (orange) with a uniform distribution (black). The results are terrible: the true probability of type I error is around $0.38$ for small nominal significance levels.

\begin{figure}
\includegraphics[scale=0.6]{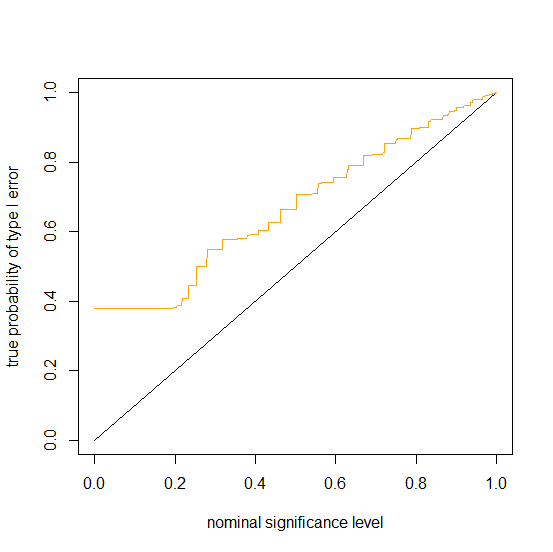}
\caption{Cumulative distribution function of $1$ minus Fisher's exact p-value (orange) versus uniform distribution (black).
$100$ patients per group, probability of ``yes'' equal to $0.05$.}\label{fig:oneminusFisher}
\centering
\end{figure}

As we will see in the next section, the situation gets worse when many p-values are combined, as even small deviations from a uniform distribution might
lead to a large deviation from a uniform distribution in the combined p-value.

\end{section}

\begin{section}{Combining p-values}
\label{sec:combining}

Given many independent p-values $\mathfrak p_1$, $\mathfrak p_2$, \dots, $\mathfrak p_n$ there are several standard techniques to combine them into a single p-value. This is often used for meta-analysis: each p-value is obtained from a test of a certain null hypothesis and the combined p-value is used to test the conjunction of all of the null hypotheses. One standard combination procedure is Stouffer's method: the combined p-value is given by:
\begin{equation}\label{eq:Stouffercomb}
\mathfrak p^{\text{Stouffer}}=\Phi\Big(\frac1{\sqrt n}\sum_{i=1}^n\Phi^{-1}(\mathfrak p_i)\Big),
\end{equation}
where $\Phi:\R\to\left]0,1\right[$ denotes the cumulative distribution function of the standard normal distribution. The rational here is the following:
if
\[\mathfrak p_i,\quad i=1,\ldots,n\]
are independent and uniform in $[0,1]$, then
\[Z_i=\Phi^{-1}(\mathfrak p_i),\quad i=1,\ldots,n\]
are independent {\em z-scores}, i.e., independent standard normal random variables. The sum $\sum_{i=1}^nZ_i$ divided by $\sqrt n$ is then again a standard normal random variable and applying $\Phi$ to a standard normal random variable we get back to a random variable with a uniform distribution in $[0,1]$, which can be used as a p-value. Stouffer's combined p-value can be thought as the p-value that arises from the test whose test statistic is given by:
\begin{equation}\label{eq:Stoufferstat}
\sum_{i=1}^n\Phi^{-1}(\mathfrak p_i),
\end{equation}
assuming that $\mathfrak p_i$, $i=1,\ldots,n$ are independent and uniform in $[0,1]$.

Another standard procedure for combining p-values is to use their product $\prod_{i=1}^n\mathfrak p_i$ or, equivalently, the sum of their logarithms
$\sum_{i=1}^n\ln(\mathfrak p_i)$ as the test statistic. If $\mathfrak p_i$, $i=1,\ldots,n$ are independent and uniform in $[0,1]$ then
$-2\ln(\mathfrak p_i)$, $i=1,\ldots,n$ are independent random variables having a chi-square distribution with $2$ degrees of freedom, so that
$-2\sum_{i=1}^n\ln(\mathfrak p_i)$ has a chi-square distribution with $2n$ degrees of freedom. In this case, the combined p-value known as {\em Fisher's combined p-value} can be computed as:
\begin{equation}\label{eq:Fishercomb}
\mathfrak p^{\text{Fisher}}=\mathbb P\Big(\chi^2(2n)\ge-2\sum_{i=1}^n\ln(\mathfrak p_i)\Big),
\end{equation}
where $\chi^2(2n)$ denotes a random variable having a chi-square distribution with $2n$ degrees of freedom.

Such combination methods work well when the individual p-values indeed have a uniform distribution, but as discussed in Section~\ref{sec:pitfalls} this is not in general the case even for ``exact'' tests. The bad news is that small deviations from uniformity can accumulate in the combined p-value. The good news is that it can be shown that if the inequalities
\begin{equation}\label{eq:inequalities}
\mathbb P_0(\mathfrak p_i\le\alpha)\le\alpha,\quad i=1,\ldots,n
\end{equation}
hold for all $\alpha\in[0,1]$ then
\begin{equation}\label{eq:combinedok}
\mathbb P_0(\mathfrak p^{\text{combined}}\le\alpha)\le\alpha
\end{equation}
also holds for all $\alpha\in[0,1]$ for reasonable\footnote{%
More precisely, it can be shown that if the statistic used for the combination method is a monotonically increasing function of the individual p-values
and if \eqref{eq:inequalities} holds then the combined p-value $\mathfrak p^{\text{true}}$ based on the true distribution of the individual p-values
is less than or equal to the combined p-value $\mathfrak p^{\text{combined}}$ obtained by pretending that the individual p-values are uniform. Since
$\mathbb P_0(\mathfrak p^{\text{true}}\le\alpha)\le\alpha$ holds, \eqref{eq:combinedok} follows.}
combination methods such as Stouffer's or Fisher's. Thus, if the individual tests are correct in the sense that true probability of type I error does not exceed nominal significance level, then the test based on the combined p-value will also be correct in the same sense. We only run the risk of the combined test being too conservative and having low power. However, as discussed in Section~\ref{sec:pitfalls}, Carlisle's method involves subtracting p-values from $1$ and in this case inequality \eqref{eq:combinedok} is reversed, leading to bad results.

Stouffer's combination method can lead to particularly absurd results when used with Fisher's exact p-values. Namely, Stouffer's combined p-value will be exactly equal to $1$ whenever any of the individual p-values is exactly equal to $1$ (assuming a continuous extension of $\Phi$ to $[-\infty,\infty]$). To avoid such absurd results, N. Brown \cite{Brown} suggests replacing individual p-values larger than $0.98$ with $0.98$. Fisher's combination method does not behave badly when an individual p-value is equal to $1$ and does not require this type of adjustment, but as we will see both Brown's adjusted Stouffer's method and Fisher's method for combining p-values lead to bad results after subtraction from $1$, even when individual p-values do not deviate much from a uniform distribution.

In Section~\ref{sec:pitfalls} we looked into the case of $100$ patients per group and a probability of ``yes'' of $0.05$ and we observed a large deviation
of Fisher's exact p-value from a uniform distribution (with similar results for chi-square test with Yates continuity correction). Now we will consider instead $100$ patients per group and a probability of ``yes'' of $0.5$, which leads to a smaller deviation between the distribution of Fisher's exact p-value and a uniform distribution (see Figure~\ref{fig:p05}). The expected value of Fisher's exact p-value in this case is equal to $0.5766$, closer to $\frac12$ than in the example considered in Section~\ref{sec:pitfalls}.

\begin{figure}
\includegraphics[scale=0.6]{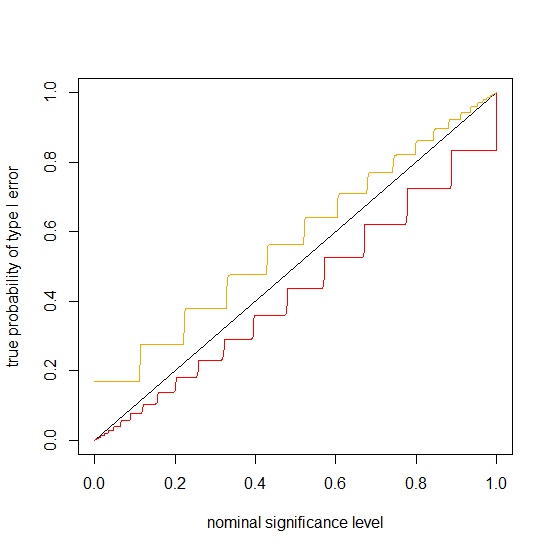}
\caption{Cumulative distribution function of Fisher's exact p-value (red) and of $1$ minus Fisher's exact p-value (orange) versus uniform distribution (black). $100$ patients per group, probability of ``yes'' equal to $0.5$.}\label{fig:p05}
\centering
\end{figure}

Let us now investigate what happens when $20$ Fisher's exact p-values are combined, assuming always $100$ patients per group and a probability of ``yes'' of $0.5$. The number of possible $20$-tuples of contingency tables is $101^{40}$, so that direct computation of the distribution of combined p-values is impossible. We thus estimate the distributions using one hundred thousand Monte Carlo simulations. Results can be seen in Figure~\ref{fig:StoufferFisher}. The curves are cumulative distribution functions of combined p-values subtracted from $1$. As expected, Stouffer's method yields absurd results (red). Brown's adjusted Stouffer's method (blue) and Fisher's method (green) are much better, but still really bad. The expected value of Stouffer's combination is approximately $0.99$, of Brown's adjusted Stouffer's combination is approximately $0.81$ and of Fisher's combination is approximately $0.72$, while the expected value of a uniform distribution is of course $\frac12$.

\begin{figure}
\includegraphics[scale=0.6]{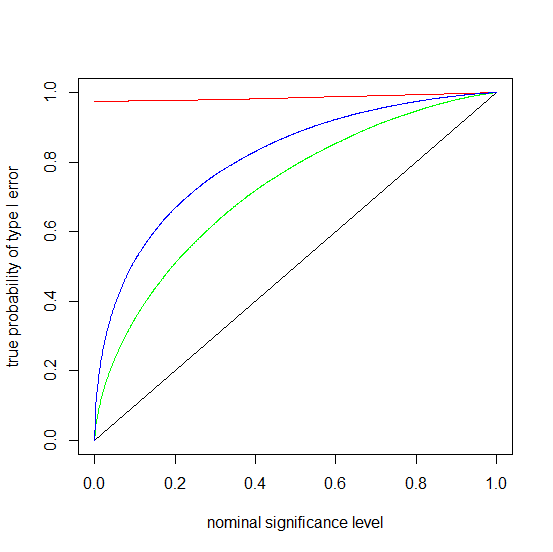}
\caption{Cumulative distribution function of $1$ minus Stouffer's combination (red), of $1$ minus Fisher's combination (green), of $1$ minus Brown's adjusted Stouffer's combination (blue) and of uniform distribution (black). $100$ patients per group, $20$ contingency tables, probability of ``yes'' equal to $0.5$. Individual p-values are Fisher's exact p-values.}\label{fig:StoufferFisher}
\centering
\end{figure}

\end{section}

\begin{section}{Adapting Carlisle's method to dichotomous variables correctly}
\label{sec:correctmethod}

As discussed in Section~\ref{sec:pitfalls}, the naive reversion of a p-value $\mathfrak p$ obtained by replacing $\mathfrak p$ with $1-\mathfrak p$ leads to problems, as inequality \eqref{eq:pbelowuniform} stating that true probability of type I error cannot exceed nominal significance level is reversed, so that true probability of type I error becomes greater than or equal to nominal significance level. Here is another way of looking at the problem: if $T$ is used as test statistic and if we define the p-value $\mathfrak p$ associated to an observed value $t$ of $T$ as $\mathfrak p=\mathbb P_0(T\le t)$ (possibly also conditioning on the value of some other random object, see Footnote~\ref{foot:conditional}) then the naively reversed p-value $1-\mathfrak p$ is equal to $\mathbb P_0(T>t)$. Thus, the naively reversed p-value excludes the observed value $t$ from the p-value. If $T$ is continuous, so that $\mathbb P_0(T=t)=0$, or at least if $\mathbb P_0(T=t)$ is very small, this does not create problems. However, for tests involving contingency tables with small numbers the probability $\mathbb P_0(T=t)$ is usually not small.

There is a simple way of fixing this problem: define the reverse p-value by $\mathbb P_0(T\ge t)$, including the observed value $t$ of $T$ as usual. With
such definition, inequality \eqref{eq:pbelowuniform} remains valid. So, for instance, we define the {\em reverse Fisher's exact p-value\/} for an observed contingency table as the hypergeometric probability of the set of all tables, with same marginals as the observed table, having a hypergeometric probability greater than {\em or equal to\/} the observed contingency table.

For the combination of the individual reverse Fisher's exact p-values one can use the usual combination formulas by Stouffer \eqref{eq:Stouffercomb} or Fisher \eqref{eq:Fishercomb}, but since the distribution of individual p-values can deviate substantially from a uniform distribution this can lead to non optimal tests with low power. The best approach is to use instead Stouffer's test statistic \eqref{eq:Stoufferstat}
or Fisher's test statistic $\sum_{i=1}^n\ln(\mathfrak p_i)$ and the exact distribution of individual p-values under the null hypothesis (conditioning as usual on the marginals of contingency tables to eliminate nuisance parameters, obtaining a hypergeometric distribution for the tables entries). Since Stouffer's statistic degenerates to $+\infty$ when any of the individual p-values is equal to $1$ and since this occurrence is not uncommon among p-values arising from contingency tables with small numbers, we suggest using Fisher's test statistic for which this inconvenience does not occur. This seems like a better option than some dirty adjustment like replacing p-values larger than $0.98$ with $0.98$ (see also Figure~\ref{fig:StoufferFisher}).

In case we have both dichotomous and continuous baseline variables, we can just combine everything by using the sum of the logarithms of all reverse p-values as test statistic, using reverse Fisher's exact p-value described above for dichotomous variables and the simple reversion $1-\mathfrak p$ for continuous variables; also, for continuous variables, we assume a uniform distribution for the p-value under the null hypothesis, which should be a reasonable approximation\footnote{%
One could run into some problems here with small samples, if the tests used to compute p-values are not exact. Also, the fact that means and standard deviations presented in articles tables are rounded could potentially create a problem, though perhaps the fact that rounding goes in both directions will create some form of compensation and alleviate this problem.\label{foot:round}}. As there is no simple closed formula for the distribution of the test statistic, one should estimate the combined p-values using Monte Carlo simulations.

\end{section}

\begin{section}{K. Sheldrick on Marik et.\ al}
\label{sec:Sheldrick}

Now let us look at a few concrete examples of incorrect applications of Carlisle's method to dichotomous variables. We start with Kyle Sheldrick's accusations against Marik et.\ al \cite{Marik}. Sheldrick's analysis was posted on his blog \cite{Sheldrick} and the original post has now been deleted for unclear reasons; one related post \cite{SheldrickCrawford} in which Sheldrick replies to critics is still online. The original post is archived here \cite{wayback} and it is still favorably cited by MedPage Today \cite{medpage}. I have left several observations in the comment section of the post explaining some of the issues discussed in this manuscript and got no response from Sheldrick. My adaptation of Carlisle's method discussed in Section~\ref{sec:correctmethod} is partially inspired by the comments of P. Shearer in Sheldrick's post (see also \cite{Shearer, Shearer2}). Sheldrick exerts no caution in his choice of words:
\begin{quote}
``{\em Unfortunately within about 5 minutes of reading the study it became overwhelmingly clear that it is indeed research fraud and the data is fabricated.

While usually I would use cautious language of `unusual' or `unexpected' patterns in the data and describe `irregularities' and `concern'; no such caution is warranted in this case. This is frankly audacious fraud. I have not requested access to the raw data or contacted the authors for explanation as the case is audacious no other explanation is apparent.}''
\end{quote}

His justification for such bold claim is based on the distribution of Fisher's exact p-values calculated for all $22$ dichotomous baseline variables
reported in Table~1 of \cite{Marik}. In Figure~\ref{fig:Sheldrickplot} we reproduce his plot of such p-values.

\begin{figure}
\includegraphics[scale=0.3]{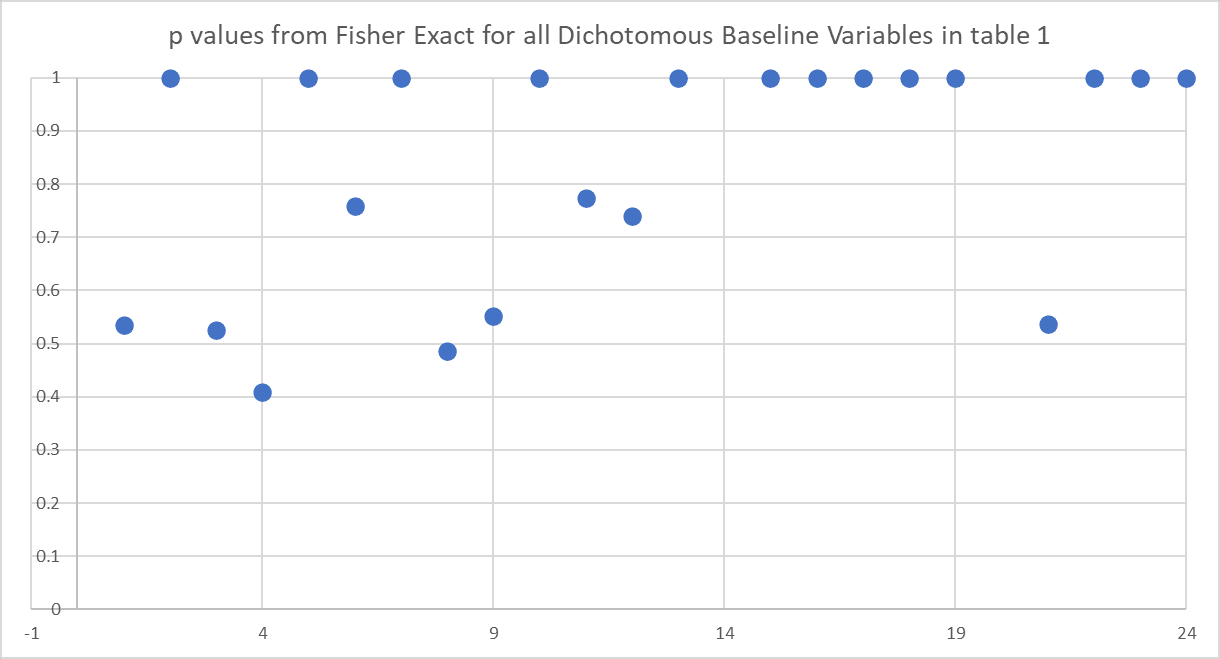}
\caption{Sheldrick's plot of all $22$ Fisher's exact p-values calculated from dichotomous variables from Table~1 of \cite{Marik}.}\label{fig:Sheldrickplot}
\centering
\end{figure}

Sheldrick apparently believes that Fisher's exact p-values for baseline dichotomous variables in a randomized trial should look like a random sample from a uniform distribution, as the following quote indicates:
\begin{quote}
``{\em Based on this we would expect that if there was no systemic bias the p values for differences in dichotomous baseline characteristics (gender, demographics, comorbidities, diagnoses etc) would, in most circumstances, centre on 0.5. If systemic differences existed however the groups may be less similar and p values may tend to numbers below 0.5, and this would not be suspicious in a non-randomised study. Systemic biases to p values greater than 0.5 are not usually possible without matching (or some very rare pseudo-block designs not relevant here) except in the setting of fraud.}''
\end{quote}
However, as discussed in Section~\ref{sec:pitfalls}, this is simply wrong. To illustrate this point in this concrete example, let us look for instance at the mean of the $22$ p-values. The mean of a random sample of size $22$ from a uniform distribution has an approximately normal distribution with mean $\frac12$ and standard deviation $\frac1{\sqrt{12\cdot22}}\cong0.062$. However, one can easily calculate explicitly the true expected value of the mean of the $22$ Fisher's exact p-values in this case (conditional on the marginals of contingency tables) and the result is approximately $0.643$. Thus, if you are expecting a random sample from a uniform distribution, the true expected value of the mean will be more than $2.3$ standard deviations above the value that you expect.

Likely following Carlisle, Sheldrick tries first to quantify the weirdness of this sample of p-values by using a Stouffer combination, but he realises that this will be a problem as Stouffer's combined p-value is equal to $1$ whenever any of the p-values being combined is equal to $1$:
\begin{quote}
``{\em This actually presents a slight problem in estimating how unlikely these results are, as the most common test for fraud in this situation would be the Stouff-Fisher, but this will declare these results infinitely unlikely (as the majority of variables have a p value of exactly 1), when in reality it is probably more likely that it is no more than trillions to quadrillions to one.''}
\end{quote}
The ``infinitely unlikely'' is an attempt to make sense of the null p-value obtained by subtracting Stouffer's combined p-value from $1$; since this is absurd, Sheldrick speculates that the correct p-value should be something like ``trillions to quadrillions to one''. But as we have discussed before, defining a p-value as $1$ minus a Stouffer combination of Fisher's exact p-values leads to disastrous results (see Figure~\ref{fig:StoufferFisher}).

His final attempt at quantifying the weirdness of the $22$ p-values is this:
\begin{quote}
``{\em A `quick and dirty' way to assess this would be to consider that the probability of any one variable having a p value over $0.4$ is $60\%$, the binomial probability of $22$ such measures having no observed values under $0.4$ is close to $(0.6)^{22}$ or around $1$ in $100{,}000$, this assumes independence which is perhaps a little unfair, but likely underestimates the improbability of such a finding as the results are not evenly distributed between $0.4$ and $1$.}''
\end{quote}
Basing a statistical test on a specific number like $0.4$ which is chosen due to properties observed in the data is not good practice\footnote{%
Formally, if you have a family $(\mathfrak p_\lambda)_{\lambda\in\Lambda}$ of random variables that are valid p-values and if you choose the index $\lambda$ by a criterion that depends on the data then $\lambda$ itself should be regarded as a random object and then you are not really using one of the valid p-values from the given family, but instead you are constructing a new random variable by plugging a random index into $\mathfrak p$. Such random variable won't in general satisfy the necessary requirements to qualify as a p-value.} and he probably realises this as he calls the method ``quick and dirty'', but another problem of course is that it is not true that we should be expecting the $22$ p-values to constitute a sample from a uniform distribution, so that the probability of an individual p-value being above $0.4$ is not $0.6$.

This concludes our discussion of Sheldrick's attempt at analysing the baseline data of \cite{Marik}. Let us now apply the correct adaptation of Carlisle's method to dichotomous variables that we suggested in Section~\ref{sec:correctmethod}. Using one million Monte Carlo simulations and the product (or, equivalently, the sum of the logarithms) of the reverse Fisher's exact p-values for the $22$ baseline variables as test statistic (and conditioning on marginals of contingency tables) an estimated p-value of approximately $1$ in $489$ is obtained.

This method of obtaining a p-value, like Carlisle's original method, assumes independence of the baseline variables, which is usually not true. In general it is not easy to do a better analysis which takes dependence into account, but in this case it is easy to do a little better than just assuming independence as $5$ of the $22$ dichotomous variables are actually the $5$ possible values of a nominal variable. More explicitly, $5$ variables are just the names of possible primary diagnoses and each patient has precisely one primary diagnosis, so that this part of Table~1 of \cite{Marik} is really a $5\times2$ contingency table. We can define a reverse Fisher's exact p-value for a given contingency table of arbitrary size by taking the probability of the set of tables, with the same marginals as the given table, whose probabilities are larger than or equal to the probability of the given table (with all probabilities conditional on the marginals of the table). In this case we have just $45{,}760$ possible tables and thus reverse Fisher's exact p-values can be calculated explicitly. The reverse Fisher's exact p-value for the $5\times2$ table corresponding to primary diagnosis in \cite{Marik} is approximately equal to $1$ in $163$. If we estimate the p-value for the other $17$ baseline dichotomous variables in \cite{Marik} using the method of Section~\ref{sec:correctmethod} (with one million Monte Carlo simulations) we get approximately $0.064$ or around $1$ in $15.5$. Thus, in some sense, the somewhat weird overbalance between groups in Table~1 of \cite{Marik} is concentrated in the $5$ lines corresponding to primary diagnosis.

Now, if we combine the p-value for the $5\times2$ table with the p-values of all other $17$ dichotomous variables by using the sum of the logarithms of all reverse Fisher's exact p-values as test statistic, we obtain a p-value of approximately $1$ in $653$ (with one million Monte Carlo simulations). This is a little less than the p-value that was obtained by treating all $22$ dichotomous variables as independent, which is counterintuitive, as one would expect that taking some dependence into account would yield a larger p-value.

Summarizing our analysis of the baseline dichotomous variables in \cite{Marik}, we conclude that p-values obtained from reasonable analyses should be of the order of $1$ in a few hundred. These are low p-values, but way above the threshold of $1$ in $10{,}000$ implicitly suggested by Carlisle \cite{Carlisle}. We note that the reported baseline data of \cite{Marik} contains also $8$ continuous variables for which mean and standard deviation are reported and there doesn't seem to be a good reason to exclude those from the analysis. In Figure~\ref{fig:ContinuousMarik} we show a plot of the $8$ p-values obtained from the continuous variables by using a t-test (assuming equal variance on both groups --- assuming unequal variance instead would make almost no difference).

\begin{figure}
\includegraphics[scale=0.6]{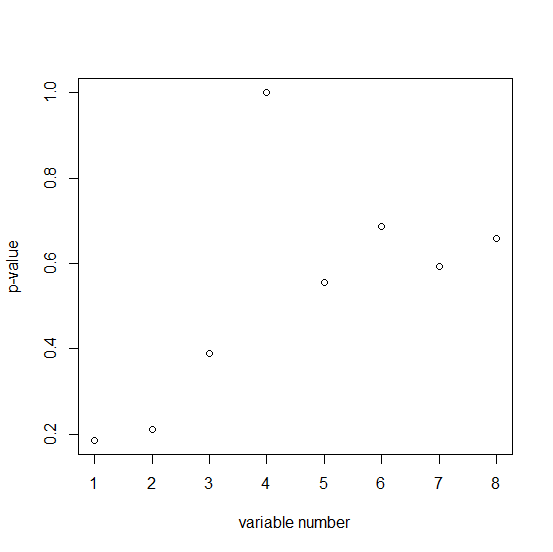}
\caption{p-values obtained from $8$ continuous variables in Table~1 of \cite{Marik} for which mean and standard deviation are reported. The p-values are computed using an unpaired t-test assuming equal variance between groups.}\label{fig:ContinuousMarik}
\centering
\end{figure}

Visually, Figure~\ref{fig:ContinuousMarik} looks like a reasonable sample from a uniform distribution. Note that a uniform distribution in $[0,1]$ has mean $\frac12$ and variance equal to $\frac1{12}\cong0.083$, while the mean of these $8$ p-values is equal to $0.535$ and the sample variance is equal to $0.072$.

We should now conclude our analysis with a formal test which includes the reverse Fisher exact p-value for the $5\times2$ contingency table corresponding to primary diagnosis, the $17$ reverse Fisher exact p-values for the other $17$ dichotomous variables and the reverse (naively, subtracting from $1$) p-values for the $8$ continuous variables for which mean and standard deviation are supplied. We use again the sum of logarithms of p-values as test statistics and assume a uniform distribution for p-values of continuous variables. Unfortunately, for one of the continuous variables (Creatinine) the reported group means in Table~1 of \cite{Marik} are exactly equal and this leads to a reverse p-value of zero, which is the case in which Fisher's statistic based on sum of logarithms degenerates to $-\infty$. Since values in Table~1 are rounded to one decimal place, we fix this problem with an ad hoc adjustment of subtracting $0.05$ from the mean of one group and adding $0.05$ to the mean of the other group for this variable (see Footnote~\ref{foot:round}).

The resulting p-value once everything is taken together is approximately $0.079$ or around $1$ in $12.7$. This is larger than even the standard ``statistical significance'' threshold of $0.05$. Of course, the fact remains that the dichotomous variables alone (or the primary diagnosis variable alone) lead to p-values of the order of $1$ in a few hundred. However, it seems to me that faking data for continuous variables with reasonable statistical properties is even harder than faking data for dichotomous variables, as one have to worry about faking also reasonable standard deviations. So, if it were the case that the patients of \cite{Marik} do not exist at all and all the data in Table~1 is fabricated, it would be a little odd that the fraudster succeeded so well with the continuous variables and not so well with the dichotomous variables. Besides a coincidence, the p-values of the order of $1$ in a few hundred for the dichotomous variables could be explained by something as simple as a mistake in the part of Table~1 related to primary diagnosis, for example.

\end{section}

\begin{section}{N. Brown on Cadegiani et.\ al}
\label{sec:Brown}

Let us now look at the application by Nick Brown \cite{Brown} of the erroneous adaptation of Carlisle's method to dichotomous variables in two studies by Cadegiani et.\ al \cite{proxa, duta}. Unlike Sheldrick (see Section~\ref{sec:Sheldrick}) Brown includes both continuous and dichotomous baseline variables in his analysis, but the number of continuous variables in either article is very small. In \cite{proxa} we have $53$ dichotomous baseline variables and only $2$ continuous baseline variables, while in \cite{duta} we have $81$ dichotomous baseline variables and only $3$ continuous baseline variables (Table~1 in both articles). Brown \cite{Brown} does several sensitivity analyses which we do not discuss in detail here. The summary of what he does is the following:
\begin{itemize}
\item compute p-values for continuous baseline variables using a t-test;
\item compute p-values for dichotomous baseline variables using either a Fisher exact test or a chi-square test with Yates continuity correction --- depending on the analysis, variables with very small values are simply excluded;
\item compute $1$ minus the Stouffer combination of all p-values obtained in the previous steps.
\end{itemize}
Also, after computing the p-values, Brown replaces any p-value above $0.98$ with $0.98$ (what we call Brown's adjustment in Section~\ref{sec:combining}) to avoid the catastrophic behaviour of Stouffer's method with p-values that are exactly equal to $1$. In the case of continuous variables, Brown worries about the fact that means and standard deviations in article tables are rounded (see Footnote~\ref{foot:round}) and he claims to have compensated for rounding in a way that maximizes the t-statistic and thus minimizes the corresponding p-value. This choice works in favor of authors, i.e., it favors the non fraud conclusion, though see Remark~\ref{rem:unico} below. This does not matter much as the number of continuous variables is small and what influences results the most are the dichotomous variables.

For the proxalutamide article \cite{proxa}, the largest p-value obtained by Brown is $0.0000337$ and for the dutasteride article \cite{duta} the largest p-value obtained by Brown is $0.00000000819$. As discussed in Sections~\ref{sec:pitfalls} and \ref{sec:combining}, the p-value obtained by subtracting from $1$ the Stouffer combination of Fisher's exact p-values can lead to really bad results, even with Brown's $0.98$-adjustment (see Figure~\ref{fig:StoufferFisher}). Moreover, replacing Fisher's exact test with chi-square test with Yates continuity correction makes little difference (see Figure~\ref{fig:Fisherchiyates}).

Applying the correct adaptation of Carlisle's method that we suggested in Section~\ref{sec:correctmethod} to the baseline data of the proxalutamide article \cite{proxa}, including both dichotomous and continuous variables, with one million Monte Carlo simulations a p-value of approximately $0.017$ or around $1$ in $60$ is obtained. Not particularly weird given the limitations of the method. I didn't worry about the fact that reported means and standard deviations in Table~1 of the articles are rounded and I simply computed the p-values for continuous variables using a t-test with the data as given in the articles (assuming equal variance on both groups). For the dutasteride article \cite{duta}, including both dichotomous and continuous variables, with one million Monte Carlo simulations a p-value of approximately $0.0024$ or around $1$ in $409$ is obtained. This is weirder, but still way above the $1$ in $10{,}000$ threshold implicitly suggested by Carlisle \cite{Carlisle}.

\begin{rem}\label{rem:unico}
For article \cite{proxa}, Brown reports explicitly all the p-values that he found. For the two continuous variables, the reported p-values are $0.431$ and $0.703$. These are the maximum possible p-values compatible with the values in Table~1 of \cite{proxa} and thus the p-values that favor the hypothesis of fraud the most. This is likely a mistake, as he claims in the text to be doing the opposite. The minimum possible p-values (using t-test with equal variance) compatible with the values in Table~1 are $0.368$ and $0.226$.
\end{rem}

\end{section}

\end{document}